\documentclass[useAMS,usenatbib,usegraphicx]{mn2e}
\usepackage{times}
\usepackage{amsmath}
\usepackage{amssymb}

\newcommand{\aj}{AJ}                   
\newcommand{\araa}{ARA\&A}             
\newcommand{\apj}{ApJ}                 
\newcommand{\apjl}{ApJ}                

\newcommand{\apjs}{ApJS}               
     
\newcommand{\apss}{Ap\&SS}             
\newcommand{\aap}{A\&A}                
  
\newcommand{\aaps}{A\&AS}              
\newcommand{\mnras}{MNRAS}             
\newcommand{\pasp}{PASP}               

\newcommand{\rxj}{RX~J0942.7$-$7726}
\newcommand{\rxjAB}{RX~J0942AB}
\newcommand{\fiftyeight}{2MASS~J0942}
\newcommand{\echa}{$\eta$~Cha}
\newcommand{\epscha}{$\epsilon$~Cha}
\newcommand{\twhydrae}{TW~Hydrae}
\newcommand{\twa}{TWA}
\newcommand{\betapic}{$\beta$~Pic}
\newcommand{\betapictoris}{$\beta$~Pictoris}
\newcommand{\scocen}{Sco-Cen}
\newcommand{\scocenfull}{Scorpius-Centaurus OB Association}
\newcommand{\lccfull}{Lower Centaurus Crux}
\newcommand{\lcc}{LCC}
\newcommand{\uclfull}{Upper Centaurus Lupus}
\newcommand{\ucl}{UCL}
\newcommand{\pms}{pre-MS}
\newcommand{\masyr}{mas~yr$^{-1}$}
\newcommand{\kms}{km~s$^{-1}$}
\newcommand{\wifes}{\textit{WiFeS}}
\newcommand{\vsini}{\ensuremath{v\sin i}}
\newcommand{\msun}{$M_{\odot}$}
\newcommand{\MLB}{MLB10}

\begin{document}
\title[An isolated pre-main sequence wide binary]{RX J0942.7$-$7726AB: an isolated pre-main sequence wide binary}
\author[S. J. Murphy, W. A. Lawson and M. S. Bessell]{Simon~J.~Murphy$^1$\thanks{Email: murphysj@mso.anu.edu.au (SJM); w.lawson@adfa.edu.au (WAL); bessell@mso.anu.edu.au (MSB)}, Warrick~A.~Lawson$^2$\footnotemark[1] and Michael~ S.~Bessell$^1$\footnotemark[1] \\
$^1$ Research School of Astronomy and Astrophysics, The Australian National University, Cotter Road, 
Weston Creek ACT 2611, Australia \\
$^2$  School of PEMS, University of New South Wales, Australian Defence Force Academy, Canberra, ACT 2600, Australia }

\maketitle
\begin{abstract}
We report the discovery of two young M-dwarfs, \rxj\ (M1) and 2MASS J09424157$-$7727130 (M4.5), that were found only 42 arcsec apart in a survey for pre-main sequence stars surrounding the open cluster $\eta$ Chamaeleontis. Both stars have congruent proper motions and near-infrared photometry. Medium-resolution spectroscopy reveals that they are coeval (age 8--12~Myr), codistant (100--150~pc) and thus almost certainly form a true wide binary with a projected separation of 4000--6000~AU.  The system appears too old and dynamically fragile to have originated in \echa\ and a traceback analysis argues for its birth in or near the \scocenfull. RX~J0942.7$-$7726AB joins a growing group of wide binaries kinematically linked to \scocen, suggesting that such fragile systems can survive the turbulent environment  of their natal molecular clouds while still being dispersed with large velocities. Conversely, the small radial velocity difference between the stars ($2.7\pm1.0$~\kms) could mean the system is unbound, a result of the coincidental ejection of two single stars with similar velocity vectors from the OB association early in its evolution. 

\end{abstract}

\section{Introduction}\label{sec:intro}

Since the discovery of a group of young stars associated with the `isolated' T Tauri star TW Hydrae \citep{de-la-Reza89,Gregorio-Hetem92,Kastner97} it has become clear that the solar neighbourhood is bestrewn with sparse associations of stars with ages substantially less than the Pleiades \citep{Zuckerman04a,Torres08}. Although members of these young associations share similar kinematics, ages and distances, their proximity to us means they are spread over vast swathes of sky. 

The youngest and best-characterised of the groups are associated with the nearby ($\lesssim$100~pc) bright stars \twhydrae\ \citep[$\sim$10~Myr;][]{Webb99}, \betapictoris\ \citep[$\sim$12~Myr;][]{Barrado-y-Navascues99,Zuckerman01}, and the neighbouring $\eta$ and $\epsilon$~Chamaeleontis \citep*[$<$8~Myr;][hereinafter \MLB]{Mamajek99,Feigelson03,Murphy10}. All four associations are located in the southern hemisphere around the \scocenfull\ (Fig.~\ref{fig:rxjskyplot}), the closest site of recent large-scale massive star formation \citep{Blaauw64,de-Zeeuw99}. 

Much work has been done to understand their origins \citep*[e.g.][]{Mamajek00,Mamajek01,Sartori03,Makarov07,Fernandez08}. Most recently, \citet{Ortega09} undertook a detailed kinematic study of the region and  concluded that the groups were likely formed in small molecular cloudlets on the outskirts of the \lccfull\ (\lcc) and \uclfull\ (\ucl) subgroups of \scocen, triggered into star formation by the bulk flows and shocks formed by colliding stellar-wind and supernovae-driven bubbles around the subgroups. With the exception of the denser \echa\ cluster, which appears to have only recently lost its natal molecular material \citep{Mamajek00}, the cloudlets later dispersed to reveal sparse, unbound associations of stars with similar ages and velocities. \citet{Fernandez08} proposed a similar scenario, with the star formation agent being the passage of a spiral density wave through the region.

Such models are reminiscent of the `in-situ' star formation process proposed by \citet{Feigelson96} to explain the discovery of isolated pre-main sequence (\pms) stars around the Chamaeleon and Taurus--Auriga dark clouds. Under this scenario, stars born in different parts of a molecular cloud inherit the region's turbulent velocity dispersion, which can be up to 10~\kms\ on scales of 10--100~pc \citep{Larson81}. In the maximal case, 10~Myr-old stars can hence be dispersed up to $\sim$100~pc away from the main cloud (1~\kms$\approx1$~pc~Myr$^{-1}$), several tens of degrees on the sky at the  distances of these complexes. Young, unbound associations such as TW Hya and \betapic\ were presumably dispersed from Sco-Cen in a similar manner, but with smaller \emph{internal} velocity dispersions (1--2~\kms), appropriate for their smaller spatial scales.

In this article we describe the discovery and characterisation of a pair of isolated, young, low-mass stars with common ages, distances and kinematics. Only 42~arcsec apart, we will show they form a wide (4000--6000~AU) $\sim$10~Myr-old binary system which may have been born in the region surrounding \scocen, near one of the young groups described above. As such wide, fragile systems are prone to disruption by $N$-body encounters, they offer insight into the dynamical conditions throughout the formation and evolution of sparse young groups and associations.

\section{\rxj\ and \fiftyeight}

\begin{figure} 
   \centering
   \includegraphics[width=0.48\textwidth]{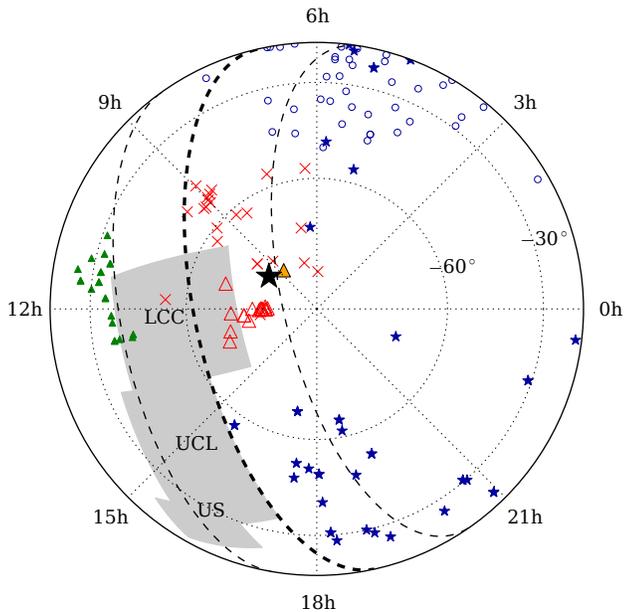}
   \caption{Orthographic projection of the southern sky, with \rxjAB\ (black star), \echa\ (orange triangle) and  members of the \epscha\ (open triangles), TW Hya (green triangles), \betapic\ (blue stars), Columba (open circles) and Carina (red crosses) Associations from \citet{Torres08}. The three subgroups of the \scocen\ OB Association \citep{de-Zeeuw99} dominate the field around the Galactic equator (thick dashed line, with $b=\pm20$\degr).}
   \label{fig:rxjskyplot}
\end{figure}

\rxj\ (=2MASS J09424962$-$7726407) and 2MASS J09424157$-$7727130 (hereinafter \fiftyeight) were identified by \MLB\ \citep[see also][]{Murphy12c} in their survey for dispersed members of the southern open cluster $\eta$ Chamaeleontis (age 5--8 Myr, $d=94$~pc). They are located 3.5 degrees to the north-east of \echa, between the cluster and the neighbouring \epscha\ Association (3--7~Myr, 90--120~pc). The \lccfull\ subgroup of the \scocen\ OB Association is immediately northward, with an estimated age of 11--17~Myr and a distance of 110--120~pc \citep{Preibisch08} (Fig.~\ref{fig:rxjskyplot}).

Although \fiftyeight\ lay within $\sim$1~mag of the empirical \echa\ isochrone and had an intermediate gravity suggestive of youth, it was rejected as a cluster member by \MLB\ due to a low Li\,{\sc i} $\lambda$6708 equivalent width compared to known members ($\sim$350~m\AA; see \S\ref{lithium}). \rxj\ was similarly excluded on account of its bad kinematic solution and colour-magnitude diagram (CMD) placement.  At the time, \MLB\ noted that \rxj\ may be an outlying member of \lcc\ but did not recognise it lay only 42~arcsec from \fiftyeight\ and possessed a similar proper motion and radial velocity (Table~\ref{table:stars}).   

\rxj\ was found to be the Weak-lined T Tauri Star (WTTS) counterpart to a \emph{ROSAT} X-ray detection by \citet{Alcala95,Alcala97} during their survey of Chamaeleon X-ray sources. They obtained low-resolution spectroscopy which confirmed the youth of the star and assigned a spectral type of K7--M0. \citet{Covino97} refined the spectral type to M0, and from a high-resolution ($R\approx20,000$) spectrum derived a stellar rotational velocity ($v\sin i \approx9$~\kms), radial velocity ($16.4\pm2$~\kms) and strong lithium absorption ($\textrm{EW}=490$~m\AA). No close companions have been found around the star, down to separations of 0.13\arcsec\ and a $K$-band contrast of $<$3.8~mag \citep{Kohler01b}. 

\fiftyeight\ is not as well characterised. Aside from survey photometry and  astrometry, and the spectroscopy presented in \MLB, the mid-M \pms\ star remains largely unstudied. 

\begin{table}
\caption{\rxj\ and \fiftyeight}
\label{table:stars}
\begin{tabular}{llll}
\hline
& \rxj\ & \fiftyeight\ & Source\footnotemark[1] \\
\hline
R. A. (J2000) &  09 42 49.62  & 09 42 41.57 & 2MASS\\
Decl. (J2000) & $-$77 26 40.8 & $-$77 27 13.0 & 2MASS\\
Spectral Type & M1 & M4.5 & this work\\
$V$ (mag) & 13.59 & --- & P06 \\
$i$ (mag) & 11.49 & 14.11 & DENIS\\
$i-J$ (mag) & 1.127 & 1.726 & 2MASS\\
$\mu_{\alpha}\cos\delta,\mu_{\delta}$ & $(-23,+17)\pm6$ & $(-16,+23)\pm9$ & PPMXL\\
 (\masyr) & $(-20,+8)\pm7$ & $(-22,+13)\pm5$ & SSA\\
& $(-25,+21)\pm(1,3)$ & --- & UCAC3\\
RV (\kms) & $20.7\pm0.4$ & $18.0\pm0.9$ & this work\\
 & $16.4\pm2$ & --- & C97\\
 & $18.5\pm0.6$ & --- & FEROS\\
\vsini\ (\kms) & $9\pm3$ & --- & C97\\
H$\alpha$ EW (\AA) & $-3$ & $-7$ & \MLB\\
Li\,{\sc i} EW (m\AA) & $450\pm50$ & $350\pm50$ & \MLB\\
& $490\pm15$ & --- & C97\\
$\log L_{\rm X}/L_{\rm bol}$ & $-2.66\pm0.18$ & --- & A97 \\
\hline
& \multicolumn{2}{c}{Weighted average} \\
\hline
RV (\kms) & \multicolumn{2}{c}{$19.9\pm0.5$} & this work\\
$\mu_{\alpha},\mu_{\delta}$ (\masyr) & \multicolumn{2}{c}{$(-21,+19)\pm(5,5)$} & PPMXL\\
\hline
\end{tabular}
(1) P06: \citet{Padgett06}, C97: \citet{Covino97}, A97: \citet{Alcala97}, \MLB: \citet{Murphy10}, SSA:  SuperCOSMOS. 
\end{table}

\begin{figure}
\centering
\includegraphics[width=0.43\textwidth]{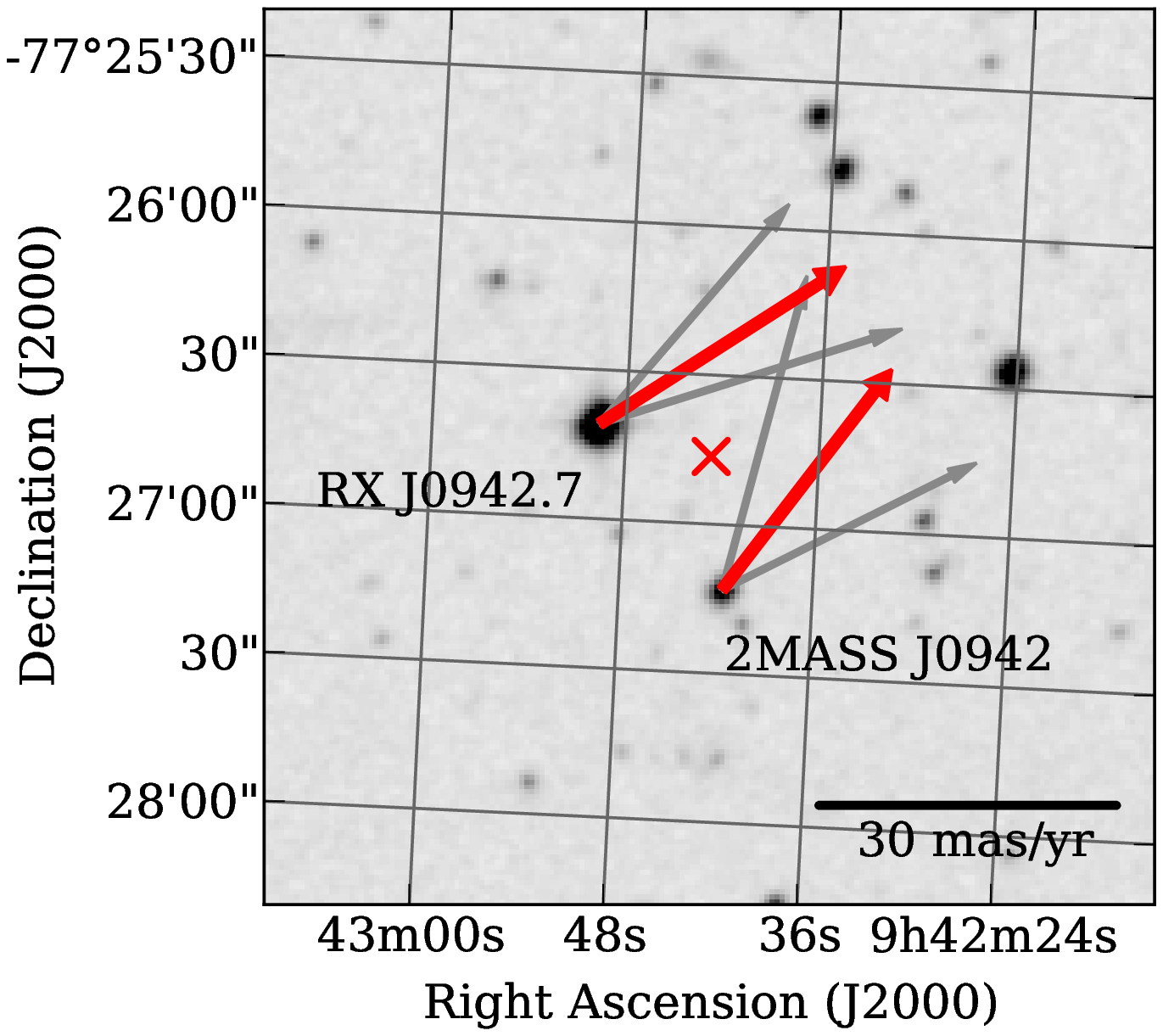} 
\caption{$3'\times3'$ DSS2-IR finder chart centred on the \emph{ROSAT} position of \rxj\ (red cross). The 1$\sigma$ uncertainty in this X-ray position is $\sim$20~arcsec, comparable to the 22~arcsec offset between it and the optical position. \fiftyeight\ was not detected by \emph{ROSAT}. The optical sources of \rxj\ and \fiftyeight\ are 42~arcsec apart. Their PPMXL proper motions are shown in red and trace the expected motion on the sky over  $\sim$2000~years.  Thin grey arrows depict the $\pm1\sigma$ error bounds.}
\label{fig:0942}
\end{figure}

Observed parameters for each star are given in Table~\ref{table:stars}.  The  radial velocities in the table are updates to the multi-epoch, medium-resolution ($R=7000$) values described by \MLB\ and are derived from 7--10 observations of each star. As orbital motions are generally negligible ($<$1~\kms) in this regime, the components of wide binaries should have kinematics that agree within errors. While their mean radial velocities only agree at the 2.7$\sigma$ level, \rxj\ and \fiftyeight\ have congruent proper motions in both the SuperCOSMOS \citep{Hambly01b} and eXtended Position and Proper Motion \citep*[PPMXL;][]{Roeser10} catalogues (\rxj\ is also found in UCAC3). We adopt the proper motions from PPMXL as it is the largest homogeneous astrometric catalogue on the International Celestial Reference System currently available and (unlike the higher-precision UCAC3) returns matches for both stars.  A Digitized Sky Survey finder chart for the system is plotted  in Fig.~\ref{fig:0942}. For the sake of brevity we hereinafter refer to it as \rxjAB. 

\section{Other stars in the vicinity}\label{sec:otherstars}

\begin{figure}
\centering
\includegraphics[width=0.485\textwidth]{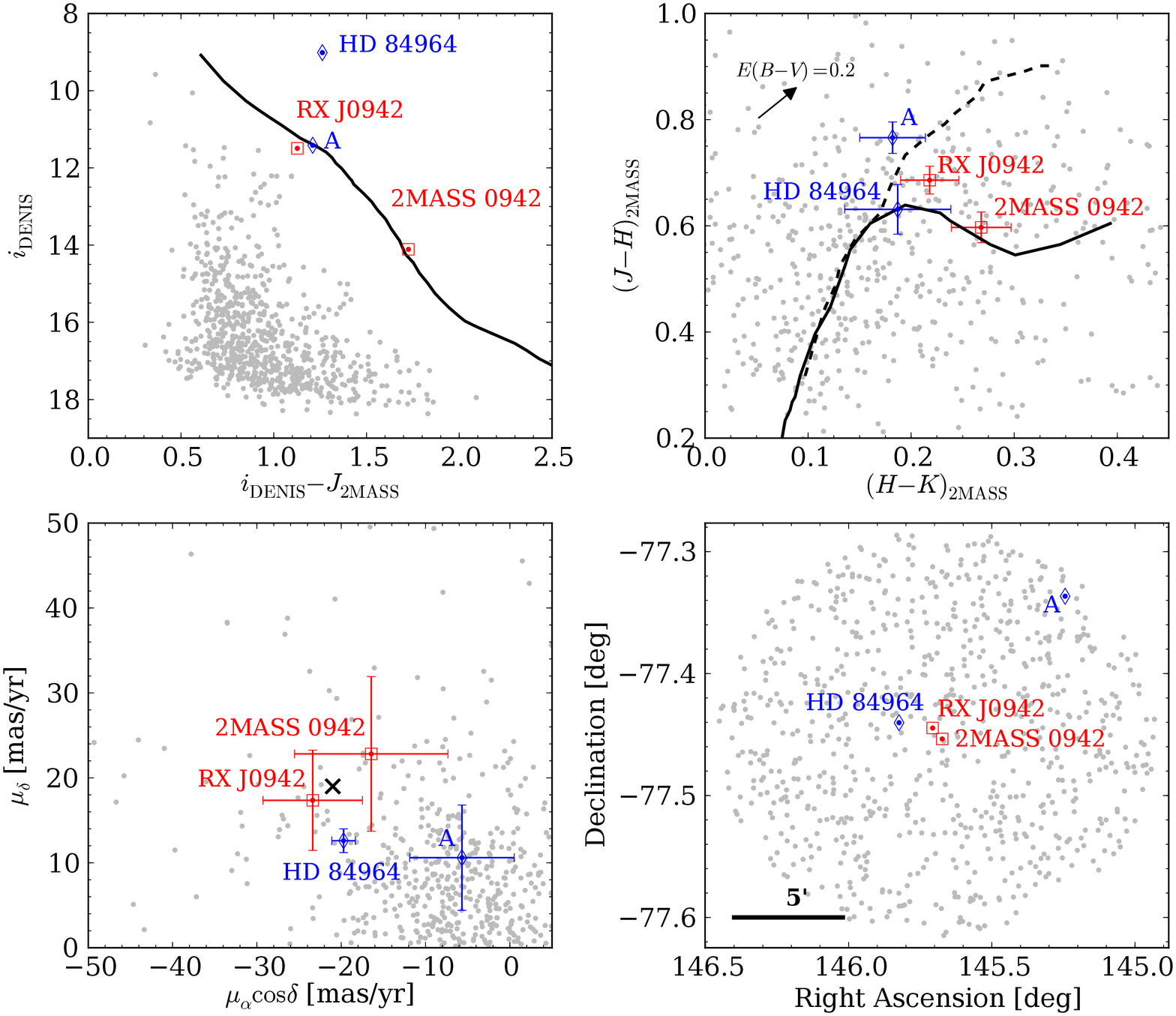} 
\caption{The 768 stars within 10\arcmin\ of \rxjAB\ with $iJHK$ photometry and PPMXL proper motions. The two blue labelled points are discussed in the text. \emph{Top-left:} Colour-magnitude diagram with a 12~Myr \citet{Baraffe98} isochrone at 150~pc for reference. \emph{Top-right:} 2MASS two-colour diagram with dwarf (solid line) and giant (dashed) locii from \citet{Bessell88}, transformed to 2MASS colours. \emph{Bottom-left:} PPMXL proper motions. The cross shows the weighted average proper motion of \rxjAB. \emph{Bottom-right:}~The~10\arcmin\ radius field on the sky.}
\label{fig:otherstars}
\end{figure}

Because \rxjAB\ lay at the bottom of the $\pm$1.5~mag selection band in the \echa\ CMD, there may be other young stars in the immediate vicinity of the pair that were missed by \MLB. Fig.~\ref{fig:otherstars} shows the 768 stars within 10~arcmin of \rxj\ with DENIS and 2MASS photometry and PPMXL proper motions. Only two stars have elevated positions in the CMD similar to \rxjAB. Star~A (2MASS~J09405860$-$7720117) lies at the edge of the 10~arcmin field but has similar photometry to \rxj. However, its 2MASS colours are giant-like and its proper motion is significantly smaller.  HD 84964 is a K1 star that \citet{Olsen94} identified as a giant from Str\"{o}mgren photometry. We note it has 2MASS colours consistent with an early K-giant and $E(B-V)\approx0.2$, the reddening attributed to the region (see \S\ref{sec:rxjred}). Although HD 84964 is only 1.5~arcmin from \rxjAB\ and has a proper motion vector near both stars, it is likely unrelated. 

To confirm this we observed HD 84964 and Star~A with \emph{WiFeS}/$R7000$ (see \S\ref{sec:obs}) on the ANU 2.3-m telescope in 2011 July and found them both to be mid-K stars (from visual comparison to radial velocity standards) with strong H$\alpha$ absorption and negligible ($<$50~m\AA) Li\,{\sc i} $\lambda$6708 equivalent widths. Their radial velocities ($67\pm2$~\kms\ and $1.4\pm2$~\kms, respectively) are also  inconsistent with either \rxj\ or \fiftyeight\ (weighted mean velocity $19.9\pm0.5$~\kms). 

We now address the likelihood that \rxjAB\ is a true wide binary, rather than the chance alignment of unrelated young stars. Given the low spatial density of \pms\ stars in the region \citep[][\MLB]{Alcala97}, it is unlikely that two such stars should lie only 42 arcsec apart and share similar kinematics. To qualify this statement we considered the $\sim$2.5$\times10^{5}$ cross-matched 2MASS/DENIS detections within 3~degrees of \rxjAB. Of the 1574 stars lying in a band $\pm$1~mag around around both stars with $(i_{\rm DENIS}-J_{\rm 2MASS})>1$, 819 had a PPMXL proper motion within 20~\masyr\ of \rxjAB. Using these 819 stars we performed a nearest neighbour analysis, after correcting for isolated pairs (where $d_{\rm AB}=d_{\rm BA}$). Of the 567 \emph{unique} nearest-neighbour distances, only four are less than 42~arcsec. This implies that the chance alignment probability of \rxjAB\ is only $P(d<42\arcsec)=4/567=0.7\pm0.4$~per cent (Poisson error). Even ignoring their obvious youth, it is thus highly unlikely \rxjAB\ is the coincidental alignment of two unrelated M-dwarfs with similar proper motions. To confirm binarity requires showing  that \rxj\ and \fiftyeight\ are coeval, codistant and have congruent space motions. 

\section{Interstellar reddening to \rxjAB}\label{sec:rxjred}

Assuming an M0 spectral type, \citet{Alcala97} used the ($V-I_{C}$) colour of \rxj\ to derive a reddening of $E(B-V)=0.27$~mag. \citet{Sartori03} found a lower value (0.17~mag) from comparison to updated model atmosphere colours, again using an M0 spectral type. The \emph{integrated} reddening in the direction of the system is $E(B-V)=0.34$~mag \citep*{Schlegel98}.

However, these are likely overestimates of the true reddening to \rxjAB. In \MLB\ we estimated the reddening towards \fiftyeight\ to be negligible from its synthetic $(R-I)_{C}$ colour.  This is consistent with the study of \citet{Knude98}, who estimated the extinction in the region as a function of distance using \emph{Hipparcos}. While \rxjAB\ lies just outside their surveyed area, they found that the reddening within 150~pc is small ($<$0.05--0.1~mag) but increases by a factor of four at the distance of the clouds \citep[160--200~pc, also see][]{Whittet97}.  Furthermore, the position of \rxjAB\ in the 2MASS two-colour diagram (Fig.~\ref{fig:otherstars}) does not support large reddenings. Comparing the photometry of \rxj\ and \fiftyeight\ to that of (unreddened) \echa\ members again suggests $E(B-V)<0.1$~mag as an upper limit on the reddening to the system.

\subsection{Observations}\label{sec:obs}

\begin{figure}
\centering
\includegraphics[width=0.48\textwidth]{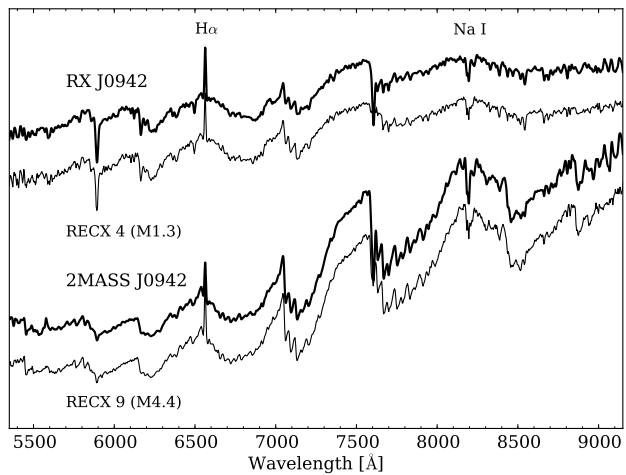}
\caption{\wifes\ $R3000$ spectra of \rxj\ and \fiftyeight. Shown for comparison are two members of \echa\ as observed by \citet{Lyo04a}; RECX~4 (M1.3) and RECX~9 (M4.4). All the spectra have been normalised at 7500~\AA. The \wifes\ spectra are smoothed to the same resolution as the \citeauthor{Lyo04a} observations.}
\label{fig:binaryspectra}
\end{figure}

To garner conclusive spectral types and robust reddenings we observed \rxj\ and \fiftyeight\ with the Wide Field Spectrograph (\wifes) on the ANU 2.3-m telescope during 2011 July. The $R3000$ grating and $RT560$ dichroic yielded a spectral resolution of $R=3000$ with wavelength coverage from 5300--9600~\AA. Because \wifes\ \citep{Dopita07} is an integral-field spectrograph, care was taken to align each star on the same image slices to ensure accurate flux calibration. The frames  were reduced using \textsc{iraf} and \textsc{figaro}\footnote{see http://www.aao.gov.au/figaro/} routines in a similar manner to that described in \citet{Murphy12c} and \citet{Riedel11}. After flat-fielding against a quartz-iodine lamp, extraction and wavelength calibration, each spectrum was corrected for telluric absorption using a contemporaneous observation of the DC white dwarf EG 131. Prior to forming the telluric spectrum, the weak H$\alpha$ and He\,{\sc i} $\lambda$5876 stellar lines were removed from the otherwise featureless white dwarf spectrum. The spectra were finally flux calibrated to remove the remaining instrumental effects using EG 131 as a flux standard. \wifes\ $R3000$ spectra of \rxj\ and \fiftyeight\ are plotted in Fig.~\ref{fig:binaryspectra}, with \echa\ members RECX 4 and 9 from \citet*{Lyo04a} for comparison. 

\subsection{\rxj}

To determine a spectral type for \rxj\ we compared its flux-calibrated spectrum to the SDSS M-dwarf template spectra of \citet{Bochanski07} and the \echa\ members observed by \citet{Lyo04a}. A visual comparison immediately showed an M0 spectral type was a poor match to the observed spectrum, irrespective of the assumed reddening. However, the M1 SDSS template and the M1.3 \echa\ member RECX~4 both gave excellent matches (see Fig.~\ref{fig:binaryspectra}). We tried de-reddening the observed spectrum but in all cases it was best fitted by the M1 templates and negligible reddening. The broad-band photometry of \citet{Alcala95} and \cite{Padgett06} also suggest little or no reddening  when compared to the average M1 colours of \citet{Bessell91}. The previous M0 spectral type gave inconsistent reddenings of $-0.02<E(B-V)<+0.17$~mag, depending on the colour index used. In light of these updated observations, we reclassify \rxj\ as an M1 star with negligible reddening.

\subsection{\fiftyeight}

The only previous spectral type for \fiftyeight\ (M4.6), was determined by \MLB\ solely from $(R-I)_{C}$ synthetic photometry. The emergence of molecular bands (e.g. TiO, VO, CaH) in the spectra of M-dwarfs enable a spectral type determination that can be effectively reddening-independent. To determine a spectral type for \fiftyeight\ we adopted a selection of molecular indices from the compilation of \citet{Riddick07} which have been shown to vary smoothly with spectral type across the mid-M subclasses (see their Table 3). These gave an average spectral type of M4.6~$\pm$~0.2. The agreement between this and the spectral type derived from  broad-band photometry immediately suggests any reddening towards \fiftyeight\ must be small. Indeed, visual comparison of \fiftyeight\ to the M4.4 \echa\ member RECX~9 \citep{Lyo04a} shows an excellent match (Fig.~\ref{fig:binaryspectra}), with an upper limit for the reddening of $E(B-V)<0.1$~mag. Similarly, \fiftyeight\  has an earlier spectral type than  ECHA~J0841.5$-$7853 (M4.7), but is later than the M4 SDSS standard, regardless of reddening. 

Finally, we computed synthetic photometry on the \wifes\ spectrum using the passbands of \citet{Bessell05}. This gave $(R-I)_{C}=1.85\pm0.03$ and a spectral type of M4.8, based on the relation of \citet{Bessell91}. The new colour agrees with that derived from the \emph{DBS} spectrum in \MLB\ ($1.80\pm0.03$).  We adopt a spectral type of M4.5 as a sensible compromise between the various estimates, with an uncertainty of $\pm$0.3 subclasses for both stars. This includes any residual reddening and flux calibration errors.

\section{Age of \rxjAB}\label{sec:rxjage}

We noted in \MLB\ that \fiftyeight\ and \rxj\ appeared to be slightly older than the immediately adjacent 5--8~Myr \echa\ cluster. This was based on their smaller than expected Li\,{\sc i} $\lambda$6708 equivalent widths and the more dwarf-like gravity of \fiftyeight\ when compared to mid-M \echa\ members. Obtaining a robust estimate of their ages (and demonstrating their coevality) is crucial in confirming the true binary nature of \rxjAB.

\subsection{Low-gravity features}

\begin{figure}
\centering
\includegraphics[width=0.48\textwidth]{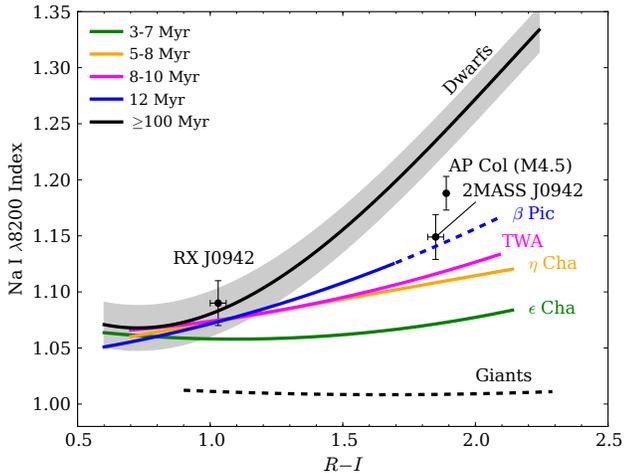}
\caption{Mean Na\,{\sc i} $\lambda$8200 indices for \rxjAB\ and members of young associations from \protect\cite{Lawson09b}, compared to field dwarfs and giants. The recently-discovered $\sim$40 Myr-old Argus member AP~Col is also plotted. \fiftyeight\ has a similar intermediate gravity, suggesting an age around 12~Myr. Errors were estimated from differences in the telluric correction, smoothing and multiple observations. The scatter around each mean trend is similar to that seen in the field dwarfs (shaded region).}
\label{fig:rxjgravity}
\end{figure}

The strength of the Na\,{\sc i} $\lambda$8183/8195 absorption doublet is highly dependent on surface gravity in mid-to-late M-type stars (\citealt{Lyo04a}; \citealt*{Slesnick06}). It can therefore be used as an age proxy for \pms\ stars contracting towards their main sequence radii. \citet*{Lawson09b} used Na\,{\sc i} doublet strengths to rank the ages of several young associations in the solar neighbourhood at a resolution of 1--2~Myr. While  there is some scatter in the gravity indices for each association, the mean trends (Fig.~\ref{fig:rxjgravity}) agree completely with the isochronal (and lithium depletion, see \S\ref{lithium}) age ranking of the groups. 

To place \rxjAB\ in Fig.~\ref{fig:rxjgravity} we computed the same Na\,{\sc i} index, $\int{F_{8148-8172}}/$ $\int{F_{8176-8200}}$, after smoothing the \wifes\ $R3000$ spectra to the $R\approx800$ resolution of the \citeauthor{Lawson09b}\ data and resampling to the same wavelength scale. As expected for an M1 star, \rxj\ is not discernible from older field dwarfs in this diagram. In constrast, \fiftyeight\ has a gravity index between dwarfs and giants, consistent with an age equal or greater than that of the \betapic\ Association \citep[10--12~Myr,][]{Torres08}.  We estimate an approximate upper age limit for \fiftyeight\ in this diagram of 40--50~Myr by comparing the star to AP Columbae, a recently identified, nearby ($8.4\pm0.1$~pc) member of the Argus Association \citep{Riedel11}. Both stars have similar spectral types ($\sim$M4.5) but AP Col has the stronger doublet strength, indicative of a slightly older age. This has been well-established as 40--50~Myr through a variety of methods, including HR-diagram placement, lithium depletion and accurate kinematics.

\subsection{Lithium measurements} \label{lithium}

\begin{figure}
\centering
\includegraphics[width=0.48\textwidth]{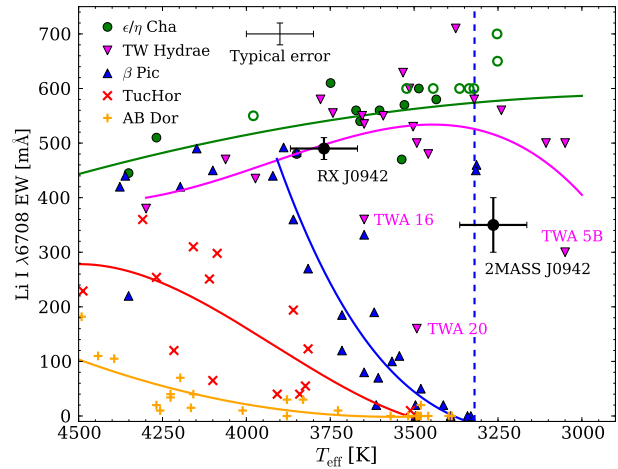}
\caption{Lithium equivalent widths of \rxjAB\ compared to other young associations from \citet{da-Silva09}; $\eta$/\epscha\ (3--8~Myr), \twhydrae\ (8--10 Myr), \betapic\ ($\sim$12 Myr), Tucana ($\sim$30 Myr) and AB~Doradus (70--120 Myr). Open symbols are the new \echa\ candidates from \MLB. Low-order polynomial fits are shown for each association. The dashed line gives the approximate location of the \betapic\ lithium depletion boundary.}
\label{fig:rxjlithium}
\end{figure}

Stars that achieve temperatures greater than $\sim$$2.5\times10^{6}$~K in their cores will start to burn lithium. The deep convection zones in low-mass stars ensure that once lithium burning has begun it will rapidly deplete the element throughout the star. Hence, the amount of lithium depletion seen in low-mass stars can serve as a mass-dependent clock over \pms\ time-scales. We plot in Fig.~\ref{fig:rxjlithium} the Li\,{\sc i} $\lambda$6708 equivalent widths of 110 K and M-dwarfs from the high-resolution study of \citet{da-Silva09}, who investigated lithium depletion in many of the recently-identified young associations reviewed by \citet{Torres08}. Equivalent widths for \rxjAB\ are plotted, with the high-resolution value of \citet{Covino97} ($490\pm15$~m\AA) used for \rxj. Following the work of \mbox{\citeauthor{da-Silva09}}, effective temperatures were estimated from $(V-I_{C})$ (\rxj) and $(R-I)_{C}$ (\fiftyeight) colours and the transformation of \citet{Kenyon95}. While improved contemporary M-dwarf temperature scales exist in the literature \citep[e.g.][]{Luhman03}, for consistency with the \citeauthor{da-Silva09} sample we adopt the \citeauthor{Kenyon95} scale. The exact choice of transformation used in Fig.~\ref{fig:rxjlithium} is unimportant as we are interested only in \emph{relative} ages. 

\rxj\ has a Li\,{\sc i}~$\lambda$6708 EW consistent with the 8--10~Myr-old TW Hya Association, some of whose late-type members have already begun to show significant depletion. We discuss the three \twa\ members with $T_{\rm eff}<3700$~K and $\textrm{EW}<400$~m\AA\ briefly below. TWA 16 is among the group of members (TWA 14--19) proposed by \citet{Zuckerman01b}  that lie south of TW Hya near HR 4796. By comparing photometric rotational periods,  \citet{Lawson05} claimed TWA 14--19 are in fact background LCC members. TWA 20 was rejected as a \twa\ member by \cite*{Song03} based on its low lithium EW. \citet{Mamajek05b} and \cite{Torres08} have computed kinematic distances to TWA 16 and 20. Compared to the 90--110~pc distance to the inner edge of \lcc, their $\sim$70~pc distance means both stars are likely true outlying \twa\ members \citep[however TWA 20 is a suspected spectroscopic binary;][]{Jayawardhana06}.  TWA 5B is a confirmed member with a distance of 45~pc \citep{Mamajek05b}. The accelerated lithium depletion seen in these three stars is likely related to surface activity or rotation \citep{King10,Jeffries06}. 

Lithium depletion is already well advanced in the $\sim$12 Myr-old \betapic\ Association, with a steep decline in EW visible down to its lithium depletion boundary (LDB, dashed line). The mass sensitivity of lithium burning means that stars less massive than this limit still retain appreciable amounts of the element. \rxj\ is not as depleted as the early-M type members of \betapic. We therefore  estimate an upper age limit of $\sim$12~Myr from Fig.~\ref{fig:rxjlithium}. Depletion ages for single stars must be interpreted with caution however, as star-to-star variations within an association (see above) can imply vastly different ages for a presumably coeval population.

Like \rxj, \fiftyeight\ also has a level of lithium depletion consistent with a \betapic\ or \twa-like age, appearing older in this diagram than $\eta$ and \epscha\ members. The star is cooler than the \betapic\ LDB so we cannot use this to constrain its age. It is however slightly warmer than the LDB of the open cluster IC 2391, which \citet*{Barrado-y-Navascues04} place at $(R-I)_{C}\approx1.90$ (spectral type $\sim$M5). The~detection of significant lithium absorption in \fiftyeight\  ($R-I=1.85$) implies it must be younger than the lithium age of IC 2391, which both \citet{Barrado-y-Navascues04} and \citet{Jeffries05} find to be $50\pm5$~Myr. Despite lithium depletion ages appearing systematically older than isochronal ages from Hertzsprung--Russell diagrams \citep*[see][among others]{Song02a,Yee10}, the relative ages still stand and \fiftyeight\ appears younger than IC 2391 \citep[main sequence turn-off age $\sim$35 Myr;][]{Barrado-y-Navascues04}.

\subsection{Activity: X-ray emission, H$\alpha$ and rotation}

Young low-mass stars are well-known to possess active coronae and chromospheres, which are manifestations of their strong magnetic fields and fast rotation rates \citep{Feigelson99}. Unfortunately for reliable age determinations, M-dwarfs have a long adolescence, with enhanced activity that can last for many Gyr \citep*[e.g.][]{Hawley96}. The diagnostics outlined below are therefore necessary but \emph{insufficient} indicators of youth.

X-ray emission from \rxj\ was identified from \emph{ROSAT} observations by \citet{Alcala95,Alcala97}. They found a saturated ratio of X-ray to bolometric luminosity typical for young, low-mass stars, $\log(L_{X}/L_{\rm bol})=-2.66\pm0.18$. Old field M-dwarfs can possess similar levels of emission, so the detection of X-rays from the star cannot constrain its age. No X-rays have yet been detected from \fiftyeight, though given the limiting flux of the \emph{ROSAT} All Sky Survey \citep*[$2\times10^{-13}$ erg~cm$^{-2}$~s$^{-1}$; ][]{Schmitt95}, in the absence of flares an M4.5 star with saturated X-ray emission would need to lie within $\sim$30~pc (10 Myr age) or $\sim$20~pc (40 Myr) to have been detected by the satellite.  

Both stars exhibited weak H$\alpha$ emission throughout the \wifes\ observations (Table~\ref{table:stars}). No emission variability was detected within the measurement uncertainties on time-scales of days to months. The low levels of emission ($|\textrm{EW}|\leq7$~\AA) are the result of quiescent chromospheric activity, not flares or accretion from a circumstellar disc, which are characterised by strong, variable emission \citep{Murphy11}. Similar levels are observed in older field dwarfs.

The rotation of a star (\vsini) may be used as a crude clock, as young stars are expected to rotate rapidly and spin-down as they age. From their echelle spectrum, \citet{Covino97} measured $\vsini=9\pm3$~\kms\ for \rxj. In a sample of 123 M-dwarfs, \citet{Browning10} found only seven stars rotating faster than their \vsini\ $\approx 2.5$~\kms\ detection limit. They further estimated  $<$10~per cent of all early-M stars are detectably rotating. The measurement of significant rotation in the \rxj\ (M1) is therefore more (albeit qualitative) evidence~of~its~youth.

 \subsection{Infrared observations}\label{sec:rxjdisk}
 
\rxj\ was observed as part of the \emph{Spitzer Space Telescope} Cores to Discs (c2d) programme \citep{Padgett06}.  No excess emission above photospheric levels was detected out to 30~\micron, which would otherwise indicate the presence of a dusty circumstellar disc. Both stars also show no excess emission to 22~\micron\ in the \emph{Wide-field Infrared Survey Explorer} All-Sky Data Release \citep{Cutri12}. This generally implies an age greater than 5--10~Myr, by which time most inner discs are observed to have dissipated \citep{Hernandez08}. However, such a relation only holds in a statistical sense (there are still strong disk excesses seen in several TW Hya and \betapic\ members for instance) and is complicated by the fact that binary disk lifetimes (especially in the presence of a wide companion) are not well-constrained  \citep[e.g.][]{Kastner12}.

\subsection{Age of the system}

The two most quantitative age indicators -- surface gravity and lithium depletion -- yield an age for both stars similar to that of the TW Hya and \betapic\ Associations, around 8--12 Myr. While the gravity index of Fig.~\ref{fig:rxjgravity} cannot constrain the age of \rxj\ (and hence coevality), the star's position in Fig.~\ref{fig:rxjlithium}  restricts its age to $\lesssim$12~Myr. The lithium data for \fiftyeight\ is also consistent with this value, which agrees with the gravity-derived age of the star. None of the other, more qualitative age indicators (X-ray and H$\alpha$ emission, measurable rotation, lack of IR excess) contradict these estimates. We therefore conclude that \rxjAB\ are a pair of coeval \pms\ stars with an approximate age of 8--12~Myr. 

\section{Distance and kinematics}\label{sec:rxjdist}

To estimate the distance to \rxjAB, we first used the theoretical \pms\ isochrones of \citet{Baraffe98} and \citet*{Siess00}, after transforming them to the observed $(i_{\rm DENIS},J_{\rm 2MASS})$ colour-magnitude space. Distances were found by minimising the square of the residuals between the isochrone and the observed photometry. For the \citet{Baraffe98} models, the estimated distances ranged from 121--250~pc, depending on the adopted age (20--5~Myr). The \cite{Siess00} isochrones gave significantly smaller distances, 99--176~pc for the same range of ages. Both stars appeared to be codistant in each set of models. In the extreme case, an age of 1~Myr (100~Myr) yielded distances of 570~pc  and 400~pc (65~pc and 35~pc), respectively. Such extreme ages are all but ruled out by the spectroscopic indicators. Furthermore, the low reddening to the system restricts its distance to within 150--200~pc (see \S\ref{sec:rxjred}). Reddening the isochrones by the maximum allowable $E(B-V)=0.1$ increased the derived distances by 5--15~per cent, depending on the models and age.

\begin{figure}
   \centering
   \includegraphics[width=0.48\textwidth]{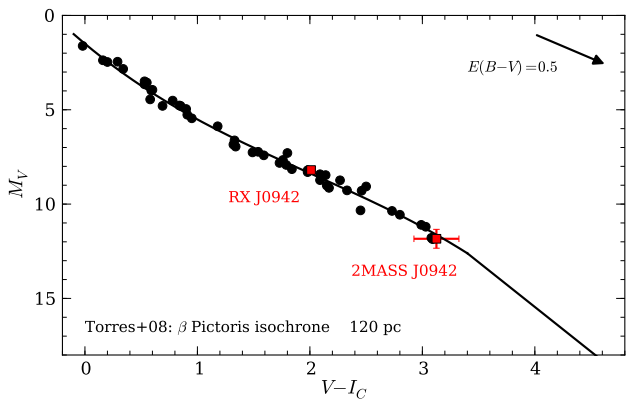} \\
   \caption{Colour-magnitude diagram for confirmed members of the \betapic\ Association with trigonometric or kinematic distances from \protect\citet{Torres08}. Photometry is taken from \protect\cite{Torres06}, \emph{Hipparcos} and DENIS. The empirical 10~Myr \betapic\ isochrone from \protect\citet{Torres08} \citep[after][]{Siess00} is also plotted.}
   \label{fig:rxjcmd}
\end{figure}

\begin{figure*}
   \centering
   \includegraphics[width=0.8\textwidth]{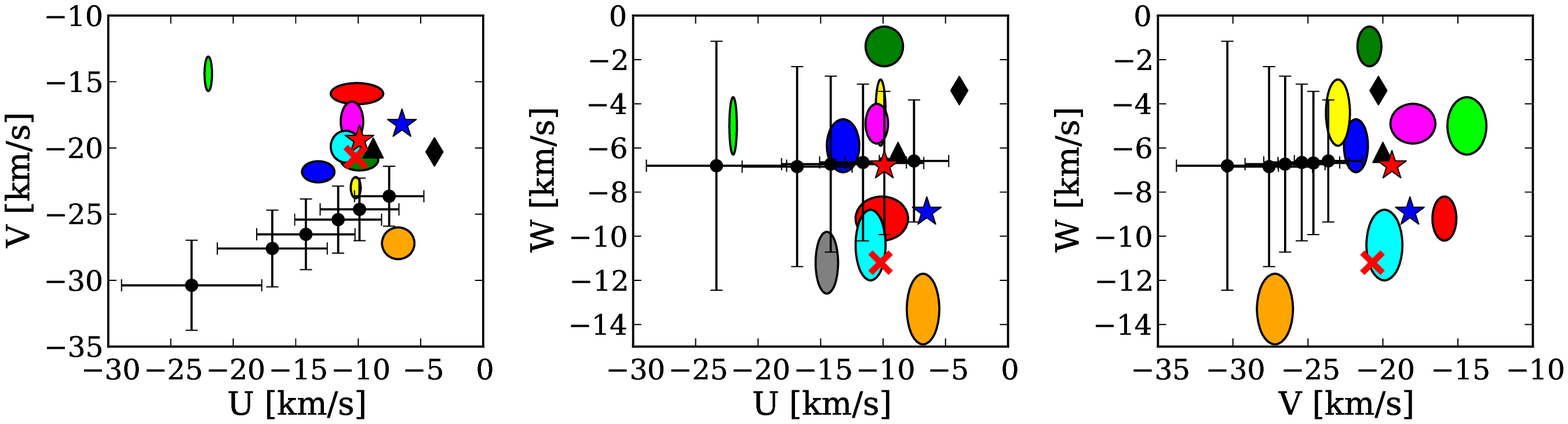}\\[-2mm]
      \includegraphics[width=0.8\textwidth]{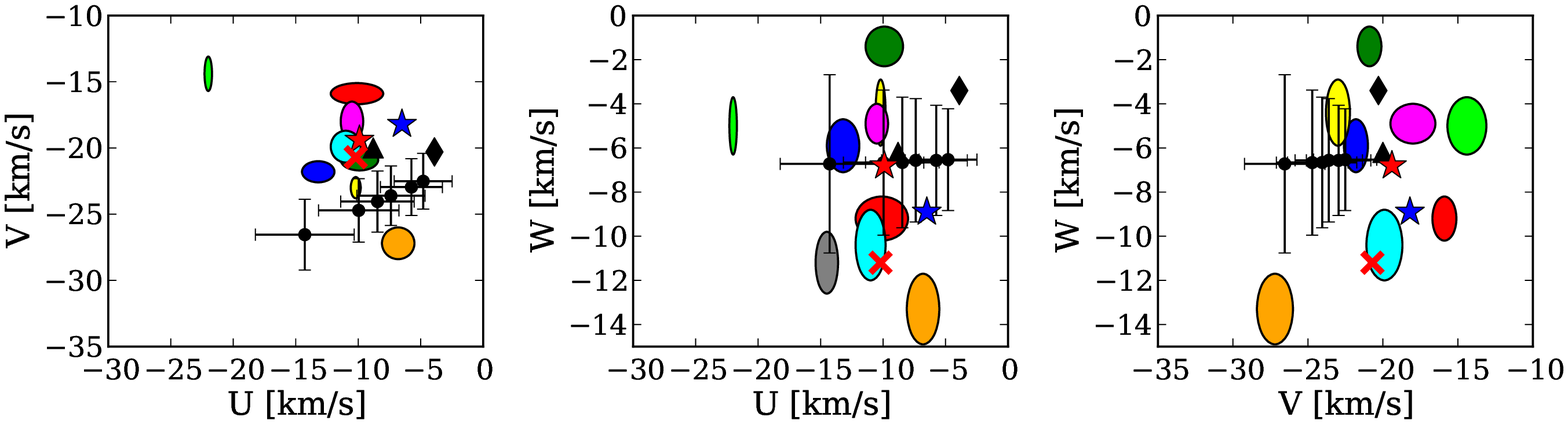} 
   \caption{$UVW$ space velocities for \rxjAB. Distances were derived from (left to right) 5, 8, 10, 12, 16 and 20~Myr \protect\citet{Baraffe98} (top) and \protect\cite{Siess00} (bottom) isochrones. Ellipses show the mean velocities and dispersions of young associations from \protect\citet{Torres08} -- Argus (light green), Columba (blue), Tuc-Hor (green), TW Hya (magenta), Carina (yellow), Octans (grey), \epscha\ (cyan), \betapic\ (red) and AB~Dor (orange). Also plotted are the mean velocities of the \lcc\ (triangle) and \ucl\ (diamond) \scocen\ subgroups, \echa\ (red cross; see \MLB), Cha~I (blue star) and Cha~II (red star). Error bars on the velocities correspond to uncertainties in the observed proper motion and radial velocity at each distance.}
   \label{fig:rxjUVW}
\end{figure*}

As a check we plot in Fig.~\ref{fig:rxjcmd} the position of \rxjAB\ in the CMD of the similarly-aged \betapic\ Association. $VI_{C}$ photometry for \rxj\ was taken from \citet{Padgett06} and for \fiftyeight\ from its $i_{\rm DENIS}$ magnitude and M4.5 spectral type, via the average colours of \citet{Bessell91}. Notwithstanding the uncertainties in transforming spectrophotometry of \fiftyeight\ to $V$ and ($V-I_{C}$) (estimated from its synthetic $R-I_{C}$ colour and 0.3 subtype spectral type error), the best-fitting distance to \rxjAB\ is 120~pc, though distances of 100--140~pc are still reasonable given the spread in the CMD.  Fig.~\ref{fig:rxjcmd} is yet more evidence \rxj\ and \fiftyeight\ are coeval and codistant. 

\subsection{Orbital motion}\label{sec:rxjbinary}

The models of \cite{Siess00} realise a mass of 0.45~\msun\ for \rxj\ and 0.2~\msun\ for \fiftyeight\ (from broad-band colours and a 10 Myr age). The \citet{Baraffe98} models give similar results but the derived masses are strongly dependent on the colour index used. We adopt the \citeauthor{Siess00} masses as they more accurately reproduce the eventual main sequence values.  If we assume the physical separation of the stars is close to their 42~arcsec angular separation, then at a distance of 100~pc the orbital period is $\sim$325,000~years and \fiftyeight\ has an orbital velocity of $\sim$0.3~\kms\ around \rxj. Statistically, the orbital semi-major axis is likely to be slightly larger than the projected separation, yielding an even longer period and smaller orbital velocity \citep{Fischer92}. Hence, a change in relative velocity of even $\sim$1~\kms\ would be enough to perturb \fiftyeight\ out of such a wide orbit. To have avoided disruption, \rxjAB\ must therefore have been born in a quiescent dynamical environment and never interacted closely with other stars over its short lifetime. 

\subsection{Space motion}

Having estimated a distance to the system, its heliocentric $UVW$ space motion can be determined. Plotted in Fig.~\ref{fig:rxjUVW} are the velocities derived for \rxjAB\ from 5--20~Myr isochronal distances and the weighted average PPMXL proper motion and \wifes\ radial velocity (Table~\ref{table:stars}). This clearly shows that, irrespective of the assumed distance (age) to the stars or model isochrones, the space motions do not agree with any of the known young kinematic groups in the solar neighbourhood. This is perhaps not surprising considering the location of the system.  Ruling out membership in any of the nearby Chamaeleon groups (\echa, \epscha, Cha~I, II) due to their younger ages, only \twa\ and \betapic\ have ages congruent with \rxjAB\ ($\sim$8--12~Myr). However, the system lies many tens of degrees on the sky from the main concentration of either group, and at a much greater distance. Given its fragile configuration and the dearth of high-density regions or  high-mass stars to power any dynamical interactions, it is implausible \rxjAB\ was ejected from either group early in its evolution.  Alternatively, \rxjAB\ has a similar age, distance and lies adjacent to the nearby \lcc\ subgroup of \scocen\  (see Fig.~\ref{fig:rxjskyplot}), whose southern extent may be as young as 11--12~Myr \citep{Preibisch08}. However, the space motion of the system in Fig.~\ref{fig:rxjUVW} also does not agree with that previously found for the subgroup \citep{Chen11}, so it is possible some sort of dispersal process may be at work.

\begin{table*}
\begin{minipage}{0.7\textwidth} 
\caption{Results of dynamical simulations for \rxjAB\ and nearby groups}
\label{table:rxjdynamics}
\begin{tabular}{llllllllll}
\hline
& \multicolumn{4}{c}{\rxjAB\ weighted average} & & \multicolumn{4}{c}{\rxj\ only} \\
\cline{2-5}\cline{7-10} \\
Association$^\dagger$ & $\Delta UVW_{\rm min}^{\ddag}$ & $v_{\rm eject}$ & $t_{\rm eject}$ & $d_{0}$ &\quad& $\Delta UVW_{\rm min}$ & $v_{\rm eject}$ & $t_{\rm eject}$ & $d_{0}$\\
& \kms & \kms & Myr & pc & & \kms & \kms & Myr & pc \\
\hline
\twhydrae & 2.5 & 7.1 & $-$13.0 & 114 & & 1.8 & 6.0 & $-$13.5 & 101\\
Columba & 1.6 & 6.8 & $-$13.75 & 108 & & 1.4 & 5.0 & $-$17.75 & 100\\
Carina & 0.6 & 4.4 & $-$8.0 & 113 & & 0.8 & 3.9 & $-$7.5 & 100\\
\lcc & 2.6 & 3.9 & $-$20.0 & 126 & & 1.6 & 3.6 & $-$20.0 & 111\\
\ucl & 0.1 & 7.4 & $-$17.25 & 130 & & 0.9 & 7.1 & $-$17.0 & 115 \\
\hline
\end{tabular} 
($\ddag$) $\Delta UVW$ is the vector magnitude of the difference  between the observed and predicted space motion at each ejection epoch and current distance. The other quantities are given at the point of minimum $\Delta UVW$.\\
($\dagger$) Only these five groups had $\Delta UVW$ less than the expected velocity error at some region in the 50--150~pc, $-20<t_{\rm eject}<0$~Myr parameter space of the simulations.
\end{minipage}
\end{table*}

To test this hypothesis and obtain kinematic distances and ejection velocities, we performed dynamical trace-back simulations, following the method described in \MLB\ and using the epicyclic approximation to Galactic dynamics formulated by \citet*{Makarov04}. We took the present-day velocities and positions of TW Hya, \betapic\ and other young groups from \citet{Torres08} and those of \lcc\ and \ucl\ from \citet{Mamajek00} and \citet{Chen11}\footnote{Unlike \echa, which can be considered a point source, the other groups have considerable spatial extent ($\sigma_{XYZ}\approx10$--40~pc). The simulations assume the mean group position and velocity are representative of the ejection point and can be used to trace its motion back in time. Ejection from another position will change the implied ejection velocity, distance and epoch.}. The simulations were performed over a range of plausible ejection times ($-20<t<0$~Myr) and current distances ($50<d_{0}<150$~pc). Because the proper motion and radial velocity of \rxj\ dominate the weighted average of the pair, we ran the simulations for both the PPMXL/\wifes\ average values and those of \rxj\ alone, using the best-available data from UCAC3 and FEROS (see \S\ref{multiplicity}). 

The results of the trace-back simulations are summarised in Table~\ref{table:rxjdynamics}. Only five of the 14 groups plotted in Figure~\ref{fig:rxjUVW} gave acceptable $\Delta UVW$ values (see notes to Table~\ref{table:rxjdynamics}). Notably, \betapic\ is ruled out as a possible birthplace for the system. Although permitted by the simulations, origin in TW~Hya requires an implausibly large ejection velocity of 6--7~\kms.  Also allowed were the 20--40~Myr-old Carina and Columba Associations \citep[subdivisions of GAYA: the Great Austral Young Association;][]{Torres08}. Although known members of Carina (almost all solar-type) overlap the position of \rxjAB\ (Fig.~\ref{fig:rxjskyplot}) and lie at similar distances ($85\pm35$~pc), both groups are much older than the 8--12 Myr we find for \rxjAB. The similarly-aged Tuc-Hor Association is also a constituent of GAYA and is clearly more lithium-depleted than \rxj\ in Fig.~\ref{fig:rxjlithium}. Moreover, the ejection epochs for Carina and Columba are well after the formation of the groups (when interactions should be rare) and require velocity changes of 4--7~\kms\ to move \rxjAB\ to its present location.

\section{Origin of \rxjAB} \label{sec:rxjorigin}

The simulations yielded kinematic distances and ejection times from \ucl\ and 
\lcc\ similar to the distances and ages of the subgroups \citep[110--140~pc, 11--17~Myr,][]{Preibisch08}. Ages at the upper end of this range are ruled out by the lithium data for \rxj, which suggests an age no older than \betapic. A~significant age spread in the subgroups could alleviate this problem; \citet*{Mamajek02} estimated that star formation in the groups ceased around 5--10~Myr ago, so it is possible some members may be younger than the quoted age. In light of the dynamical simulations, which suggest \scocen\ could have been complicit in the formation of \rxjAB, we have identified two likely mechanisms for the origin of the system:

\subsection{`In-situ' formation in a turbulent gas flow}

The `ejection' velocities required by the simulations (4--7~\kms) are typical of the bulk turbulent or thermal motions in molecular clouds \citep{Larson81}. We speculate that \rxjAB\ was born in such a turbulent flow associated with  the subgroups during their early evolution. If the flow was laminar over scales of several thousand~AU it could presumably impart a large velocity to the system whilst keeping it intact. This scenario is analogous to those proposed for the birth of \twa, \betapic\ and \epscha\ at around the same time \citep{Ortega09,Fernandez08} and closely resembles the `in-situ' star formation of \citet{Feigelson96}, in which stars born in different parts of a molecular cloud inherit the region's turbulent velocity spread and disperse into the field over several Myr. 

Such an isolated wide \pms\ binary is not unprecedented---\citet{Feigelson06} reported the discovery of the $\sim$10~Myr, 2000~AU F0/M0 binary 51 Eri/GJ 3305. With a distance of 30~pc, they ascribed the pair to the \betapic\ Association. The system lies some 100~deg distant on the sky  (110~pc in space) from \lcc, the supposed birthplace of \betapic. 51 Eri/GJ 3305 would need to have been born in gas displaced by $\sim$10~\kms\ from the \lcc\ group velocity to move to its current location in 10--12~Myr. We have proposed a similar scenario for the birth of \rxjAB. These two systems join a growing number of similar binaries in the young groups kinematically linked to \scocen, with several members of the \twa, \betapic\ and \epscha\ Associations hosting wide ($a>1000$~AU), often hierarchical, multiple systems \citep[e.g.][]{Kastner12,Looper10b,Caballero09}. Clearly it seems that the formation of sparse stellar groups can proceed in dynamically quiescent conditions necessary for wide binary survival while still imparting modest velocity dispersions to group members.

As noted in \S\ref{sec:intro}, many of the these apparently `isolated' \pms\ stars were later found to be members of sparse associations,  spanning tens to hundreds of square degrees on the sky. While we ruled out the presence of other young stars in the immediate vicinity of \rxjAB\ in \S\ref{sec:otherstars}, it is possible \rxj\ and \fiftyeight\ are the first members of a hitherto-unknown group to be discovered. There are several hundred PPMXL detections within a 10 degree radius of the pair that have congruent photometry, distances and space motions (assuming a realistic range of radial velocities and a 10 Myr age), a handful of which lie within an arcminute  of a \emph{ROSAT} X-ray source (a first-order youth indicator). Confirmation of these stars as a coeval, co-moving association is outside the scope of this work, but it is possible that \rxjAB, like 51 Eri/GJ 3305, formed in a small, unbound group on the outskirts of \scocen\ approximately 10 Myr ago.

\subsection{Coincidental ejection from \scocen?}

Alternatively, \citet{Moeckel10} and \citet{Kouwenhoven10} found that their $N$-body simulations could produce wide ($a>10^{3-5}$ AU) binary systems in the haloes of dynamically evolving young clusters when two stars were coincidently ejected with similar velocity vectors and become weakly mutually bound. Moreover, \citet{Moeckel11} recently showed that a small, transient population of wide binaries can persist in the halo of an evolving cluster and be `frozen out' into the field when the time-scale of binary destruction exceeds that of the decreasing stellar density. Old, wide binaries are rare \citep[only a few per cent of systems have $a>10^{4}$~AU in the solar-type sample of][]{Duquennoy91}, but a weakly-bound system could explain why \rxj\ and \fiftyeight\ have radial velocities that differ by 2.7~\kms. Given the errors on the mean velocities, this is a 2.7$\sigma$ difference and well above the 0.3~\kms\ variation expected due to orbital motion. 

The clusters in the $N$-body simulations described above are typically very rich. For instance, \citet{Moeckel10} evolved a cluster of $\sim$1200 stars (stellar mass $\sim$200~\msun) with a half-mass radius of only 10$^{4}$~AU (0.05~pc). There is no observational evidence that \betapic\ or TW Hya were \emph{ever} in such a rich, dense configuration. In contrast, dynamical models show that \echa\ may have been initially very dense \citep*[10$^{8}$~stars~pc$^{-3}$;][]{Moraux07}, but likely only possessed $<$100 members at birth and is spectroscopically older than \rxjAB. In lieu of a nearby rich cluster, the early dynamical evolution of the \scocen\ OB Association---perhaps via the dissolution of non-hierarchical few-body ($N<10$) systems \citep[e.g.][]{Sterzik95}---appears to be the best scenario for the creation of \rxjAB\ from the ejection and transient, weak interaction of two single stars.

Improved kinematics (ideally from parallaxes) and better knowledge of the dispersed low-mass population of \scocen\ are needed to ultimately determine the origin of \rxjAB.  High-resolution radial velocities will make it possible to discriminate between the coincidental ejection of two low-mass stars from \scocen\ or the turbulent formation of \rxjAB\ therein.

\section{Further multiplicity?}\label{multiplicity}

Our mean \wifes\ radial velocity for \rxj\ ($20.7\pm0.4$~\kms, Table~\ref{table:stars}), differs significantly from that reported by \citet{Covino97} ($16.4\pm2$~\kms, circa 1995). The variation is well outside that expected from orbital motion around \fiftyeight\ and so may be evidence of unresolved binarity. Since a large fraction of M-dwarfs are seen in multiple systems \citep[e.g.][]{Fischer92}, it is not unreasonable to expect that many of the primaries of wide binaries are in fact close multiple systems. Recent surveys (e.g. \citealt*{Makarov08}; \citealt{Faherty10}) have confirmed this and it is also seen in $N$-body cluster simulations including primordial binaries \citep{Kouwenhoven10}. Moreover, there are hints that the frequency of hierarchical systems may increase with separation, at least for M-dwarf systems \citep{Law10}.

\cite{Kohler01b} did not detect a companion around \rxj\ in their speckle interferometry and direct imaging survey of Chamaeleon \emph{ROSAT} sources. Their 0.13\arcsec\ (3.81~mag) detection threshold, the singly-peaked cross-correlation function reported by \citet{Covino97} and the good fit to the CMD in Fig.~\ref{fig:rxjcmd} put strong constraints on the orbit and mass of any companion. For instance, a 0.1~\msun\ star in a 5~AU (0.05\arcsec\ at 100~pc) orbit would  induce an orbital velocity in \rxj\ of only 1.7~\kms\ with a period of $\sim$14 years. Such a companion could explain the velocity difference without being observed directly.

\rxj\ was observed 19 times during 1999--2005 with FEROS on the ESO-1.5m and 2.2m telescopes at La~Silla\footnote{ESO programmes 075.C-0399, 073.C-0355, 072.A-9012, (PI: E.\ Covino)}. We have obtained the reduced spectra (E. Guenther, private communication) and derive a mean velocity of $18.5\pm0.6$~\kms. Velocities were computed by cross-correlation over the region 5900--6500~\AA\ against the M1V star HD 36395, using an archival ELODIE \citep{Prugniel04} spectrum and its observed velocity \citep*[$7.6\pm0.5$~\kms,][]{White07}. No evidence of a companion was visible in the cross-correlation functions and the individual velocities show no trend over six years of observations. Some scatter outside the instrumental errors is present, probably as a result of chromospheric activity \citep{Murphy11}.

Intriguingly, this FEROS mean velocity bisects the \citet{Covino97}\ and our 2010--2011 \wifes\ values. It also agrees within errors with the velocity derived for \fiftyeight. While at first this may indicate that the two stars are co-moving, the contemporary \wifes\ velocity for \rxj\ ($20.7\pm0.4$~\kms) is inconsistent with both values. It is possible we are seeing temporal variation in the radial velocity between 1995, 1999--2005 and 2010--2011. The long period and small amplitude of the variation---if it exists at all---means further radial velocity monitoring of \rxj\ (and high-contrast imaging) is necessary to confirm any unresolved binarity.

 \section{Conclusion}
 
From new low- and medium-resolution spectroscopy we have shown that \rxj\ and \fiftyeight\ are a pair of isolated, coeval (age $\sim$10~Myr) and \mbox{codistant} (100--150~pc)  \pms\ stars deep in the southern sky. Their 42~arcsec separation and the sparsity of other young stars in their vicinity mean they almost certainly form a true wide binary with a projected separation of 4000--6000~AU. Both stars have proper motions that agree within errors and similar radial velocities. Their exact origin remains  uncertain, but we propose they were born in turbulent gas near the \lcc\ or \ucl\ subgroups of the \scocen\ OB Association. Similar birthplaces have been proposed for other wide \pms\ binaries like 51 Eri/GJ 3305 and the unbound \betapic, TW~Hya and $\epsilon$ Cha Associations, whose members were probably born in similar turbulent flows in the region over the past 5--15~Myr. 

Alternatively, the small but significant radial velocity difference (2--3~\kms) we observed could imply the system is unbound, possibly as a result of the coincidental ejection of two single stars from the subgroups with similar velocity vectors. If confirmed as bound, the existence of \rxjAB, 51 Eri/GJ 3305 and other wide binaries kinematically linked to \scocen\ suggests that such fragile systems can survive the turbulent environment of their natal molecular clouds while still being dispersed with large velocities.

\section*{Acknowledgments}

We thank Eike Guenther for providing reduced FEROS spectra of \rxj, Pavel Kroupa, Eric Mamajek and Eric Feigelson for their considered comments on the thesis of SJM from which this work is based, and the anonymous referee for their thorough review of the manuscript. This research has made use of the VizieR and SIMBAD services provided by CDS, Strasbourg and the \textsc{topcat} catalogue and VO tool, developed by Mark Taylor and available for download at http://www.star.bris.ac.uk/\textasciitilde mbt/topcat/.

\end{document}